\documentclass[photonics,article,accept,pdftex,moreauthors]{Definitions/mdpi} 

\firstpage{1} 
\pubvolume{1}
\issuenum{1}
\articlenumber{0}
\pubyear{2024}
\copyrightyear{2024}
\externaleditor{\textls[-25]{Academic Editor: Firstname Lastname}}
\datereceived{28 September 2024} 
\daterevised{25 November 2024} 
\dateaccepted{26 November 2024} 
\datepublished{ } 
\hreflink{https://doi.org/} 

\Title{Spin--Phonon Relaxation Dynamics from a Conical Intersection of Trapped Rydberg Ions}

\TitleCitation{Spin--Phonon Relaxation Dynamics from a Conical Intersection of Trapped Rydberg Ions}

\Author{Manish Chaudhary 
 $^{1,}$*
$^{,\dagger}$\orcidA{}, Rejish Nath $^{1}$\orcidB{} and Weibin Li $^{2}$\orcidC{}}

\AuthorNames{Manish Chaudhary, Rejish Nath and Weibin Li}

\AuthorCitation{Chaudhary, M.; 
 Nath, R.; Li, W.}

\address{%
$^{1}$ \quad Department of Physics, Indian Institute of Science Education and Research, Pune 411008, India; rejish@iiserpune.ac.in\\
$^{2}$ \quad School of Physics and Astronomy, and 
 Centre
 for the Mathematics and Theoretical Physics of Quantum Non-Equilibrium Systems,  University of Nottingham, NG7 2RD, 
 UK; weibin.li@nottingham.ac.uk}

\corres{Correspondence: xmanish@uliege.be}

\firstnote{Current address: Department of Physics, CESAM, University of Liège, Liège B-4000, Belgium.}  

\abstract{Non-adiabatic processes near conical intersections  are rooted in the stronger coupling between electronic and nuclear degrees of freedom.  Using a system of two trapped Rydberg ions, their high polarizability and strong dipolar interactions allow to form a conical intersection, where dynamics takes place on a microsecond time scale. Rydberg lifetimes are typically from a few to tens of microseconds, which could affect the conical dynamics. We study the effect of the finite lifetime of the Rydberg state on the vibronic dynamics around the conical intersection via analyzing the master equation. Through mean field and numerical calculations, damping dynamics are found in both the phonon populations and electronic states depending on the initial states. It is found that oscillatory vibronic dynamics can be seen clearly within the Rydberg lifetime, permitting to observe the conical effect in the trapped Rydberg ion system.}

\keyword{conical intersection; trapped Rydberg ion; master equation; dissipative process 
} 

\begin{document}




\section{Introduction}

Non-adiabatic couplings between electronic and nuclear motions~\cite{domcke2004conical} influence a wide range of chemical processes~\cite{malhado2014non}. In~the vicinity of a conical intersection (CI), two or more molecular energy surfaces degenerate at certain spatial coordinates~\cite{yarkonyDiabolicalConicalIntersections1996,juanes-marcosTheoreticalStudyGeometric2005,patersonConicalIntersectionsPerspective2005}. Due to the degeneracy,  Born--Oppenheimer (BO) approximation fails, leading to strong vibronic coupling between electronic and vibrational states~\cite{PhysRevLett.109.173201}. This has immense consequences for excited state dynamics~\cite{hauseDynamicsConicalIntersections2008} when molecular wavepackets traverse the intersection, producing, for~example, ultrafast radiationless reactions~\cite{ismail2002ultrafast,domcke2012role}. Geometrical phases associated with the quantum interference of the wavepacket around CIs are generated that slow down the nuclear motion~\cite{ryabinkin2017geometric,whitlow2023quantum,valahu2023direct}.
Investigating the dynamics around CIs serves important applications in atomic physics~\cite{PhysRevLett.103.083201,PhysRevLett.106.153002}, chemistry~\cite{domcke2012role,chen2019mapping}, solid state materials~\cite{nelson2020non}, and  biological processes~\cite{barbatti2010relaxation,polli2010conical,hammarstrom2008coupled}. Non-adiabatic dynamics in molecules typically take place on a femtosecond time scale, which requires ultra-fast and broadband spectroscopy techniques to reveal details in the vicinity of a CI~\cite{mcfarland2014ultrafast,young2018roadmap,adachi2019probing,kowalewski2015stimulated,kitney2014two}. A~growing effort has been spent to quantum simulate CIs with controllable physical systems~\cite{macdonellAnalogQuantumSimulation}, including superconductors~\cite{PhysRevX.13.011008}, ultra-cold atoms~\cite{PhysRevLett.103.083201,PhysRevLett.106.153002,brown2022direct}, trapped ions~\cite{ivanovSimulationJahnTeller2013,porrasQuantumSimulationCooperative2012,whitlow2023quantum,valahu2023direct}, and Rydberg atoms~\cite{PhysRevLett.126.233404}, to~understand and control the dynamics near CIs. Compared to real molecules,  CI dynamics occurs on a much slower time scale in these quantum simulators. The~internal and external states of the trapped ion quantum simulator can be controlled~\cite{blatt_quantum_2012,wang_simulating_2024}, which allows to monitor the dynamics and understand quantum effects in~detail.

Rydberg ions have large state-dependent polarizability~\cite{liElectronicallyExcitedCold2012,higginsHighlyPolarizableRydberg2019a,pawlakRydbergSpectrumSingle2020a,niederlanderRydbergIonsCoherent2023a,zhang2020submicrosecond} and strong dipole--dipole interactions~\cite{mullerTrappedRydbergIons2008}. In~a linear Paul trap, Rydberg ions interact with the electric field of the trap, leading to sizable electron--vibration couplings~\cite{PhysRevLett.132.223401}. Trapped Rydberg ions have emerged as a versatile quantum simulation and computation platform~\cite{mullerTrappedRydbergIons2008,higginsSingleStrontiumRydberg2017} for realizing fast quantum gates~\cite{zhang2020submicrosecond}, and~simulating novel physics, such as Dicke states~\cite{gambettaExploringNonequilibriumPhases2019}, flywheel~\cite{martinsRydbergionFlywheelQuantum2023}, spin model~\cite{nathHexagonalPlaquetteSpin2015}, non-Hermitian dynamics~\cite{lourencoNonHermitianDynamicsMathcal2022}, and~the tripartite Rabi model~\cite{hamlynTripartiteQuantumRabi2024}. Recently,  it has been proposed that a CI can be created between a pair of trapped Rydberg ions~\cite{PhysRevLett.126.233404} by merging the high degree of control over the internal and external states of  trapped ions with lasers~\cite{leibfriedQuantumDynamicsSingle2003,haffnerQuantumComputingTrapped2008}, state-dependent polarizability~\cite{zhang2020submicrosecond,PhysRevLett.119.220501}, and~tunable dipole--dipole interactions in Rydberg states~\cite{PhysRevLett.125.133602,gambettaExploringNonequilibriumPhases2019, mokhberiChapterFourTrapped2020}. This setting enables the study of the geometric effects and vibronic dynamics near a CI where the electronic and vibrational states are encoded in two Rydberg states and the breathing mode of the ion crystal, respectively. By~preparing the wavepacket in one of the potential minima around the CI, it has been shown that the wavepacket is largely localized in the initial potential minima due to the geometric phase-induced destructive interference~\cite{ryabinkinGeometricPhaseEffects2017}. 

On the other hand, dissipation plays an important role~\cite{chenDissipativeDynamicsConical2016} in, for~example, the damping of emission~\cite{gelmanDissipativeDynamicsSystem2005},~charge transfer~\cite{burghardtExcitedStateChargeTransfer2006}, and~system--bath coupling~\cite{cederbaumShortTimeDynamicsConical2005} around CIs. In~a recent experiment with a superconducting circuit simulator~\cite{PhysRevX.13.011008}, it was shown that the dephasing of electronic states enhances wavepacket branching when passing through an engineered CI. In~Rydberg states, lifetimes are long (compared to the laser--ion coupling and dipolar interaction) but finite~\cite{saffmanQuantumInformationRydberg2010,shaoRydbergSuperatomsArtificial2024a,mokhberiChapterFourTrapped2020}. How finite lifetimes of the Rydberg states will affect vibronic dynamics around the CI has not been yet investigated. In~this work, we provide a case study of the dissipative spin and phonon dynamics around the CI by taking into account the spontaneous decay in Rydberg states of Sr$^+$ ions. Lifetimes in $\vert nS_{1/2}\rangle$ states are much shorter than that of $\vert nP_{1/2}\rangle$ states ($n$ is the principal quantum number). Without~the spontaneous decay, the spin and phonon dynamics are affected by the CI; for~example, non-adiabatic effects due to the spin--phonon coupling take place. 
Based on a quantum master equation, we numerically simulate the spin--phonon dynamics with different coupling strengths and initial states. We show that  phonon populations and spin--phonon correlations oscillate initially and reach steady states at later stages depending on the~lifetime. 

The structure of the paper is as follows. 
In Section~\ref{sec2}, we introduce the physical system, energy levels and model Hamiltonian. We discuss lifetimes in the Rydberg state that are the source of dissipation in our case. In~Section~\ref{sec2}B, 
 we observe the coherent dynamics encoded in physical quantities such as the phonon number and population of both Rydberg levels without considering the dissipation.
In Section~\ref{sec:dissipation}, we introduce the dissipation and discuss the stationary state due to the spin--phonon coupling and decay using the mean field theory. Then, we solve the exact master equation numerically and analyze the time evolution of the coupled spin--phonon modes  and compare it with the mean field results.
Finally, in~Section~\ref{sec5}, we conclude the results and provide the future~perspective.


\section{Model and Coherent~Dynamics}
\label{sec2}
\unskip
%
%
%
\subsection{Hamiltonian}
\begin{figure}[H]%
\includegraphics[width=.95\linewidth]{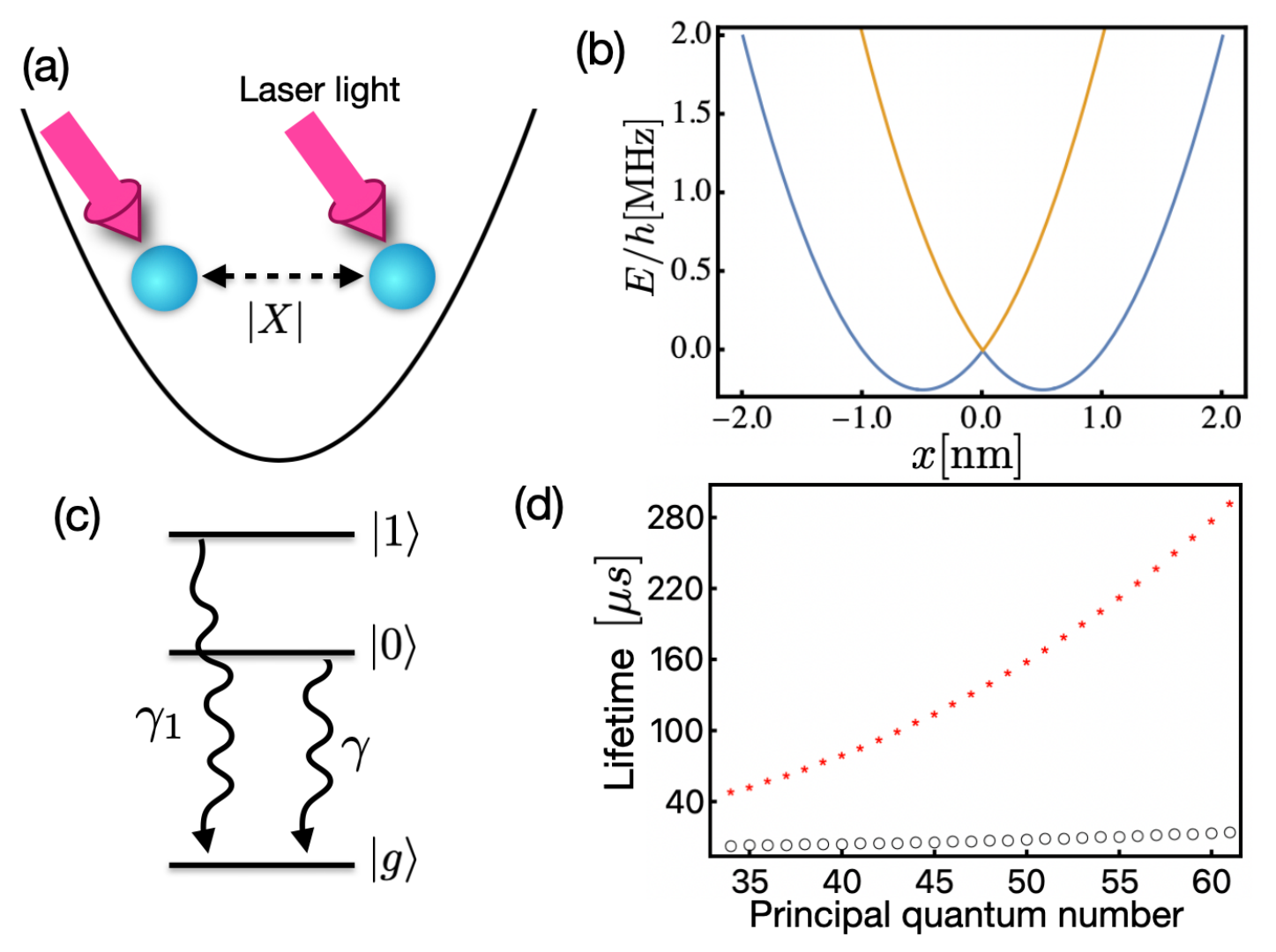}
	\caption{(\textbf{a}) Two Rydberg ions in Rydberg states $\lvert 0\rangle=\lvert nS_{1/2}\rangle$ and $\lvert 1\rangle= \lvert n'P_{1/2}\rangle$ vibrate around their equilibrium distance $|X|$ in a linear Paul trap. In~Rydberg states, the~long-range dipole--dipole interaction couples to the breathing mode of the crystal vibration. (\textbf{b}) The conical intersection is formed due to the coupling of Rydberg states $\lvert 0\rangle$ and $\lvert 1\rangle$ with the breathing mode. (\textbf{c}) Energy levels. Both Rydberg states decay to low-lying electronic states. The~low-lying state does not participate in the CI dynamics. (\textbf{d}) Lifetime of Sr$^+$ ions in the $\lvert nP_{1/2}\rangle$ state (star) and~in the $\lvert nS_{1/2}\rangle$ state (empty circle). The~lifetime in $S$ states is much shorter than that in the $P$ states.}
	\label{fig0}%
\end{figure}

Following the scheme in our previous work~\cite{PhysRevLett.126.233404}, we engineer a CI in a system of two Sr$^+$ ions in a linear Paul trap. The~setting  is depicted in Figure~\ref{fig0}a. The~ions form a two-ion crystal and vibrate around their equilibrium positions~\cite{jamesQuantumDynamicsCold1998}. Both ions are laser excited to Rydberg states coherently~\cite{higginsSingleStrontiumRydberg2017}.  One ion is in Rydberg energy level $\vert 0\rangle=\vert nS\rangle$ and the other is in the $\vert 1\rangle = \vert nP\rangle$ state, where $n$ is the principal quantum number. In~Rydberg states, the~ions interact strongly via a long-range dipole--dipole interaction. Due to the spatial dependence, the~dipole--dipole interaction between Rydberg ions couples to the breathing mode of the ion crystal~\cite{jamesQuantumDynamicsCold1998}. Using an additional static electric field and exploiting the state-dependent polarizability of Rybderg states, we create two linear couplings perpendicular to each other, forming a peaked (symmetric) CI as~shown in Figure~\ref{fig0}b. More details on the derivation of the CI can be found in Ref.~\cite{PhysRevLett.126.233404}. The atomic properties of Sr$^+$ ions are based on a model potential calculation~\cite{aymarMultichannelRydbergSpectroscopy1996}


Around the CI of the trapped ions, the~vibronic dynamics is governed by the Hamiltonian~\cite{PhysRevLett.126.233404}. We use the units with $\hbar\equiv 1$:
\begin{equation} 
 H = \frac{p_x^2}{2m} + \frac{p_y^2}{2m} + \frac{m \omega_x^2 x^2}{2} + \frac{m \omega_y^2 y^2}{2} + g_x x S_z + g_y y S_x
 \label{interham}
\end{equation}
where $m$ is the mass, and~$p_{\xi}$ and $\omega_{\xi}$  are the momentum operator and trapping frequency along the $\xi$-axis ($\xi=x,y,z$). $g_x$ and $g_y$ represent the coupling between the spin and the motional degrees of freedom. The~coupling strength can be individually tuned by varying the dipole--dipole interaction between the Rydberg ions, or~an external electric field. Operators $S_{\xi}$ are two-ion collective spin operators as
\begin{eqnarray}
	S_x = & |10\rangle \langle 01| + |01\rangle \langle 10| = \sigma^{l}_{10} \sigma^r_{01} + \sigma^l_{01} \sigma^r_{10} ,\nonumber \\ 
	S_y = & i (|10\rangle \langle 01| - |01\rangle \langle 10|) = i (\sigma^{l}_{10} \sigma^r_{01} - \sigma^l_{01} \sigma^r_{10}), \nonumber \\ 
	S_z = & |10\rangle \langle 10| - |01\rangle \langle 01| = \sigma^{l}_{11} \sigma^r_{00} - \sigma^l_{00} \sigma^r_{11}
\end{eqnarray}
where the individual projection operator $\sigma^{l,r}_{ij} = |i\rangle \langle j |$ with $i,j = g, 0,1$. In~a previous work~\cite{PhysRevLett.126.233404}, we studied the CI dynamics with the Hamiltonian~(\ref{interham}) numerically without quantizing the spatial coordinates. In~this work, we quantize the vibrations and use phonon Fock states~\cite{haffnerQuantumComputingTrapped2008}. As~a result, the Hamiltonian (\ref{interham}) can be rewritten as
\begin{eqnarray}
	H   =&  \omega_x \Big(a^{\dagger}_x a_x + \frac{1}{2}\Big) + \omega_y \Big(a^{\dagger}_y a_y+ \frac{1}{2}\Big)  
	+   G_x (a^{\dagger}_x + a_x) S_z + G_y (a^{\dagger}_y + a_y) S_x 
	\label{interham_oper}
\end{eqnarray}
where $a_\xi(a^{\dagger}_\xi)$ represents the annihilation (creation) operator of the $\xi^{\text{th}}$ phonon mode, i.e, $\xi=\eta_{\xi} (a^{\dagger}_\xi + a_\xi)$ where $\eta_\xi=\sqrt{1/2m\omega_\xi}$ is the Lamb--Dicke parameter. The~coupling parameters $G_\xi = \eta_\xi g_\xi$, and $\xi\in(x,y)$.

\subsection{Quantum Vibronic~Dynamics}
We first discuss the coherent dynamics of the system governed by the Hamiltonian~(\ref{interham_oper}), i.e.,~in the absence of spontaneous decay. It will help us to understand the effect of spin--phonon coupling and~also later compare with that of the dissipative dynamics. 
We assume that the phonons along the $x$-direction are in a coherent state, whereas those along the $y$-axis are in the vacuum state (zero phonon state). The~ion on the left-hand side is prepared in state $\vert 0\rangle$ and the right ion is in state $\vert 1\rangle$. Explicitly, the~initial state is described by
\begin{equation}
	|\psi_i\rangle = | \alpha_x \alpha_y\rangle \otimes |0, 1\rangle, \label{eq:initial_state}
\end{equation}
where the coherent state is defined in terms of phonon Fock states as~\cite{monroeSchrodingerCatSuperposition1996,scullyQuantumOptics1997},
\begin{eqnarray}
	| \alpha \rangle  = e^{-\frac{|\alpha|^2}{2}} \sum_{n = 0}^{\infty} \frac{\alpha^n}{\sqrt{n!}} |n\rangle,
\end{eqnarray}
where $\vert n\rangle$ is the phonon Fock (number) state with $n$ phonons. The~phonon number operator and its mean value are determined by $N_\xi =  a^{\dagger}_\xi a_{\xi}$, and~$\langle N_\xi\rangle = \langle a^{\dagger}_\xi a_\xi \rangle$, respectively.
%

%

\begin{figure}[H]%
\includegraphics[width=\linewidth]{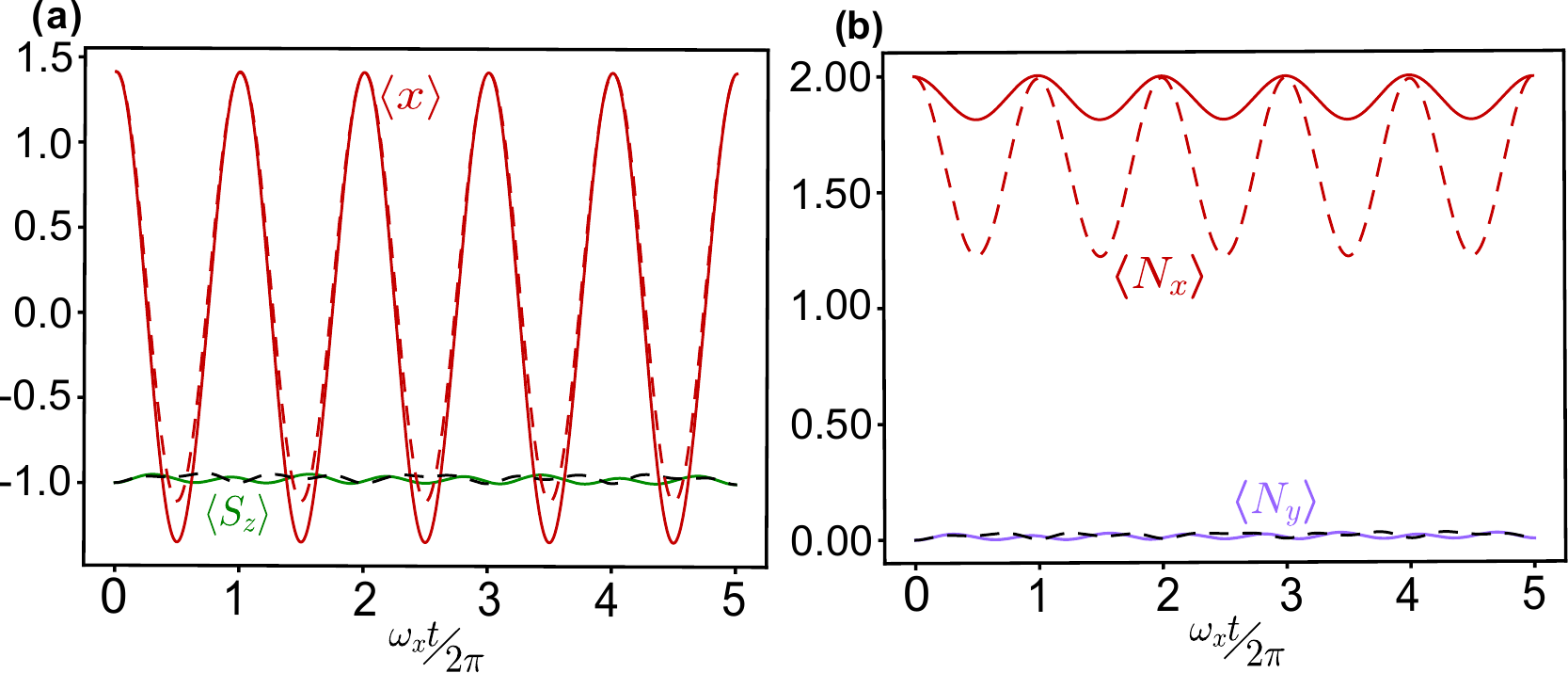}
	\caption{Coherent quantum 
 dynamics. Time evolution of (\textbf{a})  average values $\langle x\rangle$ and $\langle S_z\rangle$, and~(\textbf{b}) average phonon number $\langle N_x\rangle$ and $\langle N_y\rangle$. These quantities are obtained by solving the Hamiltonian (\ref{interham_oper}).  In~both figures, we consider different coupling $(G_x,G_y)=2\pi\times (0.22,0.86)$ MHz (solid line), and~ $(G_x,G_y)=2\pi\times (1,1)$ MHz  (dashed line), respectively. Other parameters are $\alpha_x=\sqrt{2}$, $\alpha_y=0$, $\omega_x=2\pi\times 1$ MHz, $\omega_y = 2\pi\times 1.6$ MHz and $\gamma_S = 0$. {The units of $\langle x\rangle$ are in $nm$.} 
 }%
	\label{fig1}%
\end{figure}

Upon solving the time-dependent Schr\"odinger equation $i\partial |\psi(t)\rangle/\partial t=H|\psi(t)\rangle$, we analyze the spin and phonon dynamics with the parameters that are relevant to the current experiment. The~numerical simulation is performed in the $\{\vert g\rangle,\, \vert 0\rangle,\, \vert 1\rangle\}$ spin basis in the tensor product with the phonon Fock basis. The average value $\langle \mathcal{O}\rangle$ of an operator $\mathcal{O}$ is then evaluated with $\langle \mathcal{O}\rangle=\langle \psi(t)\vert\mathcal{O}\vert \psi(t)\rangle$. Considering the trap parameters used in Ref.~\cite{PhysRevLett.126.233404}, we obtain $G_x=2\pi\times 0.22 $ MHz, and~$G_y=2\pi\times 0.86$ MHz, $\omega_x=2\pi\times 1$ MHz, and~$\omega_y = 2\pi\times 1.6$ MHz. We note that coupling $G_\xi$ can be increased, for~example, by~increasing the principal quantum number~\cite{hamlynTripartiteQuantumRabi2024}.  We see that the spin population $ \langle S_z \rangle =\langle \psi(t)|S_z|\psi(t)\rangle$ is not significantly affected as~shown in  Figure~\ref{fig1}a. The~expectation values $ \langle S_x \rangle$ and $ \langle S_y \rangle$ vanish, as~the coupling generates the spin configuration which is orthogonal to the initial~state.


Due to the spin--vibronic coupling, particularly $G_x$ and the choice of the initial state, $\langle x\rangle$ oscillates around $x_m = 2\eta_x\alpha_x = 21.44 $ nm 
while $\langle y\rangle$ remains zero. This means that the phonon wavepacket shows a localization 
around the initial location but~not a tunnel to the other minimum at $-x_m$, despite the energies at the two symmetric minima being degenerate.
This results from the geometric effect, where the phonon wavepacket will gain a $\pi$ phase shift when encircling the conical intersection. This phase leads to destructive interference such that the wavepacket at the opposite minimum vanishes~\cite{ryabinkin2017geometric,PhysRevLett.126.233404}. When the coupling is weak, there is no significant change in the phonon number in both the $x$ and $y$ directions as~shown in Figure~\ref{fig1}b. Increasing the coupling  $G_x$ results in a larger amplitude of oscillation of $\langle x\rangle$. 
Here, the~variation in the expectation value $\langle S_z\rangle$ does not change significantly with the change in the coupling and exhibits small oscillation around $-1$ . This value is close to the initial state. 
The phonon fluctuations $\langle N_y\rangle$ are relatively small, whose value is related to $\langle S_z\rangle$.  This is due to the parity symmetry, i.e.,~$\mathcal{P}H\mathcal{P}^\dagger = H$, where $\mathcal{P}=S_ze^{i\pi N_y}$, is the parity operator~\cite{braakIntegrabilityRabiModel2011,xieQuantumRabiModel2017}. Hence, ($\langle S_z\rangle + \langle N_y \rangle$) is a conserved quantity. See Appendix \ref{App A} for details. 

\begin{figure}[H]%
	\includegraphics[width=.98\linewidth]{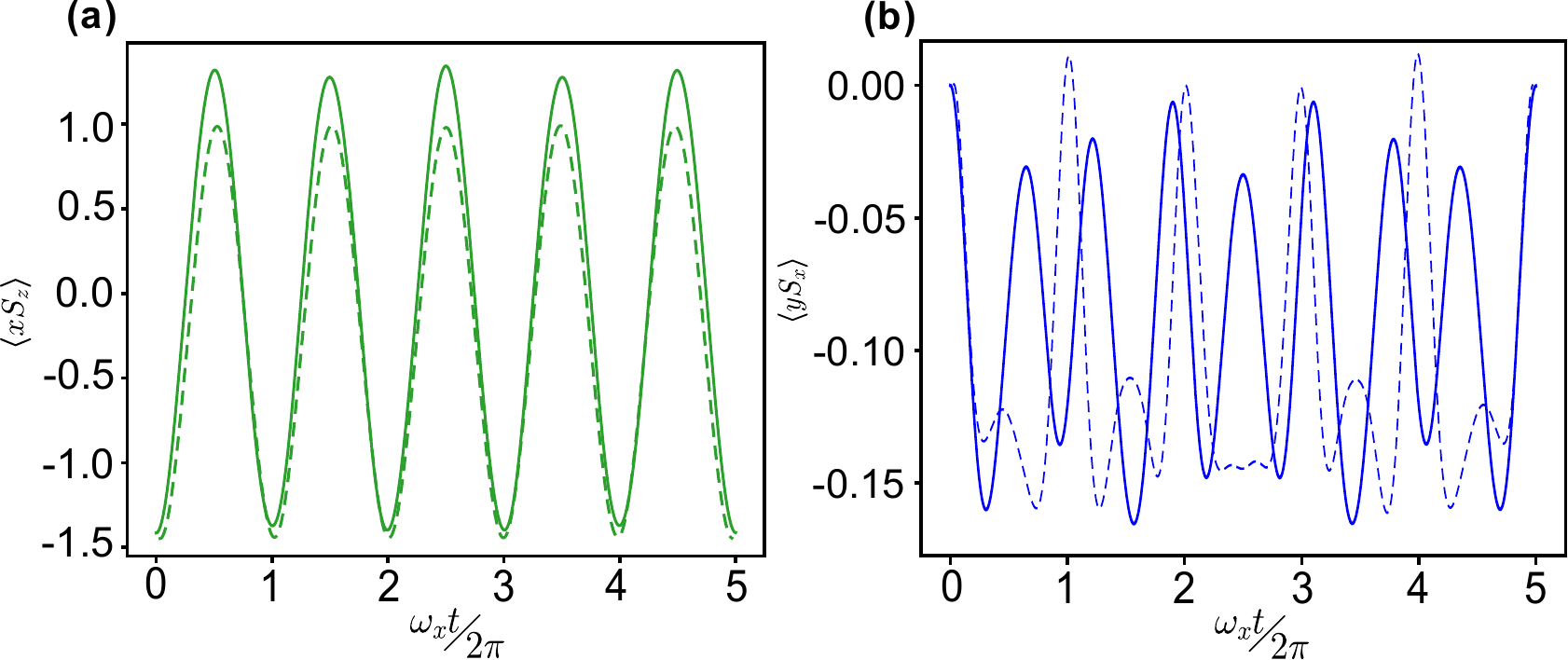}
	\caption{Correlation 
 between spin and phonon modes. We show (\textbf{a}) $\langle xS_z\rangle$ and (\textbf{b}) $\langle yS_x\rangle$. 
 Here, $(G_x,G_y)=2\pi\times (0.22,0.86)$ MHz (solid line), and~ $(G_x,G_y)=2\pi\times (1,1)$ MHz (dashed line), respectively. Other parameters are $\alpha_x=\sqrt{2}$, $\alpha_y=0$, $\omega_x=2\pi\times 1$ MHz, $\omega_y = 2\pi\times 1.6$ MHz and $\gamma_S = 0$.   \label{fig3}}
	\end{figure}

One can find that the spin and phonon dynamics are strongly correlated. Figure~\ref{fig3} shows correlations between the spin operators corresponding to Rydberg states and vibrational modes. The~spin operator $S_z$ only couples with the $x$ phonon. One finds that the expectation value $\langle x S_z\rangle$ oscillates as~shown in Figure~\ref{fig3}a. Average values  of the spin operators $S_x,S_y$ with the $y$ phonon are non-zero (Figure~\ref{fig3}b) due to the coupling $G_y\neq 0$. Correlations  $\langle x S_z\rangle$ and $\langle y S_{x}\rangle$ become stronger by increasing the coupling $G_x$ and $G_y$ as~shown in Figure~\ref{fig3}a,b.  We note that other correlation terms, such as $\langle x S_{x}\rangle$, $\langle x S_y\rangle$, and~$\langle y S_{z}\rangle $  are zero,  as~the coupling does not generate the correlations between the spins and phonons in these~directions. 

\section{Dissipative Vibronic~Dynamics}
\label{sec:dissipation}
With the illustration of the coherent CI dynamics, we now turn to the investigation of the dissipative vibronic dynamics. Rydberg states have finite lifetimes and decay spontaneously to the ground state via many intermediate states~\cite{gallagherRydbergAtoms2005}. Decay from the two Rydberg states is our main concern, while we are not interested in details of the decay via the intermediate states. With~this consideration, we will employ a model in which Rydberg states decay to the ground state, and~the intermediate steps will be neglected. The~level scheme is given in Figure~\ref{fig0}c. State $\vert g\rangle$ is a low-lying state, where Rydberg states decay directly or through cascade processes to the ground state. 
To be concrete, we will focus on Sr$^+$ ions in the~simulation.

\subsection{Rydberg Lifetime and Master~Equation}
Lifetimes in $nS$ states are a few  to a few tens of microseconds, while being much longer (tens to hundreds of microseconds) in $nP$ states when $n>30$ as~depicted in Figure~\ref{fig0}d. For~example, we find lifetimes are $7.2\,\upmu$s and $158\,\upmu$s in $50S$ and $50P$ states of Sr$^+$ ions, respectively. Lifetimes in $nP$ states are one order of magnitude larger than $nS$ states.  The~difference between the lifetimes of these two states can be turned even larger by increasing $n$. Trap frequencies and ion--phonon coupling are in the order of megahertz~\cite{PhysRevLett.126.233404}.  As~we discussed earlier, one could obtain $G_x=2\pi\times 0.22 $ MHz, and~$G_y=2\pi\times 0.86$ MHz, $\omega_x=2\pi\times 1$ MHz, and~$\omega_y = 2\pi\times 1.6$ MHz in typically linear Paul traps. As~a result, decay from Rydberg $nS$ states will play immediate roles once the dynamics starts, whose time scale is comparable to that of the trap frequency and spin--phonon coupling. Decay from the $nP$ state takes place at a much later stage, where the conical intersection is not relevant, due to the decay in the $nS$ state (i.e., the spin sector). 
In the following analysis, we will focus on the decay in the Rydberg $nS$ state (i.e., state $\vert 0\rangle$). 

Taking into account the spontaneous decay of Rydberg states,  the dynamics of the system are governed by a master equation~\cite{scullyQuantumOptics1997,breuerTheoryOpenQuantum2007},
\begin{eqnarray}
	\frac{\partial \rho}{\partial t} = -i [H,\rho] + \mathcal{D}(\rho)
	\label{mastereq}
\end{eqnarray}
where $\rho$ is the density matrix of the spin and phonon states.  The~spontaneous decay from Rydberg states $|0\rangle$ to the ground state $|g\rangle$ is described by Liouvillian operator $\mathcal{D}(\rho)$,
\begin{eqnarray}
	\mathcal{D}(\rho) = \sum_{j=1,2} \gamma_S (\sigma^{j}_{g0} \rho \sigma^{j}_{0g} - \{ \sigma^{j}_{0g} \sigma^{j}_{g0}, \rho \}/2), 
	\label{dissipator}
\end{eqnarray}
and $\gamma_{S}$ represents the decay rate for the dissipation process. Our approach differs from the work in Ref.~\cite{PhysRevX.13.011008} in the sense that we employ the controllable state-dependent polarizability and stronger dipole interactions in Rydberg states as opposed to the superconducting circuits. In~the latter case, the~decoherence is due to the dephasing. 
We now analyze the time evolution of the dissipative dynamics by solving the master equation numerically using the Qutip toolbox~\cite{qutip}. 
We analyze time evolution of the phonon and spin dynamics for the various coupling strengths and a given initial~state.

%


%
%

%
%
\subsection{Dynamics of the Collective Spin and~Phonons}
\begin{figure}[H]%
    \includegraphics[width=.98\linewidth]{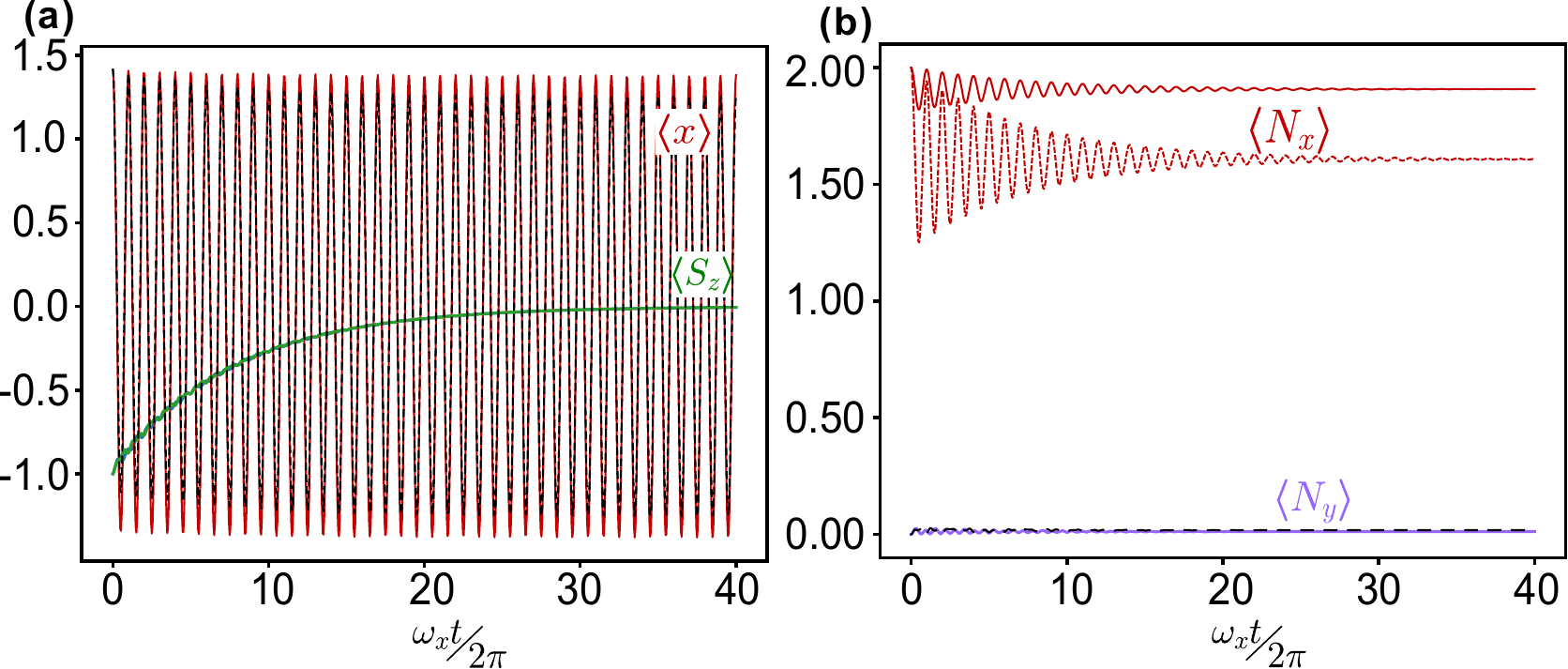}
    \caption{
    (\textbf{a}) Average values 
 of $\langle x\rangle$ and $\langle S_z\rangle$. 
    (\textbf{b}) Average phonon number $\langle N_x\rangle$ and $\langle N_y\rangle$.  Couplings are $(G_x,G_y)=2\pi\times (0.22,0.86)$ MHz (solid line), and~ $(G_x,G_y)=2\pi\times (1,1)$ MHz  (dashed line), respectively, with the dissipation parameter $\gamma_S = 0.13 \mu s^{-1}$ for the Rydberg state $50 S$. 
    Other parameters are $\alpha_x=\sqrt{2}$, $\alpha_y=0$, $\omega_x=2\pi\times 1$ MHz, and~$\omega_y = 2\pi\times 1.6$ MHz. }%
    \label{fig5}%
\end{figure}

Figure~\ref{fig5} shows the dynamical evolution of the collective spins and phonons with the initial state (\ref{eq:initial_state}) in the presence of dissipation with different coupling parameters. 
With the coupling  $G_x= 2\pi\times 0.22 $ MHz  and $G_y= 2\pi\times 0.86$ MHz, it is evident from Figure~\ref{fig5}a that the non-zero expectation value of the spin operator $|\langle S_z \rangle|$ decreases with time and attains a steady state with $\langle S_z \rangle= 0$. This is expected, as eventually, state $\vert 0\rangle$ will decay to $\vert g\rangle$ at a later time. Note that $\langle S_z \rangle= 0$ is not caused by equal populations in states $\vert 1,0\rangle$ and $\vert 0,1\rangle$ but~by the fact that the collective spin sector is destroyed, i.e.,~spin state $\vert 0,1\rangle$ is eliminated. In~other words, the expectation values of the collective spin are always zero.  This is also confirmed through our mean field calculation (see Appendix \ref{App B}).  The~mean position$\langle x\rangle$ of the $x$ phonon oscillates with time, rapidly.  Note that the oscillation of $\langle x\rangle$ persists, though~$\langle S_z\rangle$ already reaches its steady~value.

An interesting effect is observed for both the average phonon number $\langle N_x \rangle$ and $\langle N_y \rangle$ as~shown in Figure~\ref{fig5}b. Though~there is no phonon loss term in the dissipative process in the master equation, the~coupling with the spins makes the average phonon number  decrease with time. It attains the steady state value $\langle N_x \rangle^{\text{steady}}$ and $\langle N_y \rangle^{\text{steady}}$ at a longer evolution time. The~parity symmetry of the Hamiltonian, on~the other hand, is broken in the master equation due to the decay in Rydberg states such that $\langle N_y\rangle=0$ in the steady state. Increasing the coupling to $G_x=G_y= 2\pi \times 1 $ MHz,  one finds larger amplitudes in the expectation value of the $x$ phonon mode that decreases with the time, as~can be seen in Figure~\ref{fig5}b. It is clear that the average phonon number in the $x$ phonon mode is larger compared to the $y$ phonon mode. This is largely due to the choice of the initial state, i.e.,~$\alpha_x> 0$ and $\alpha_y=0$. Initially, the $y$ phonon is in the ground (zero phonon) state, which is hardly affected by the spin--phonon~coupling.
\vspace{-12pt}

%

\begin{figure}[H]%
	\includegraphics[width=.98\linewidth]{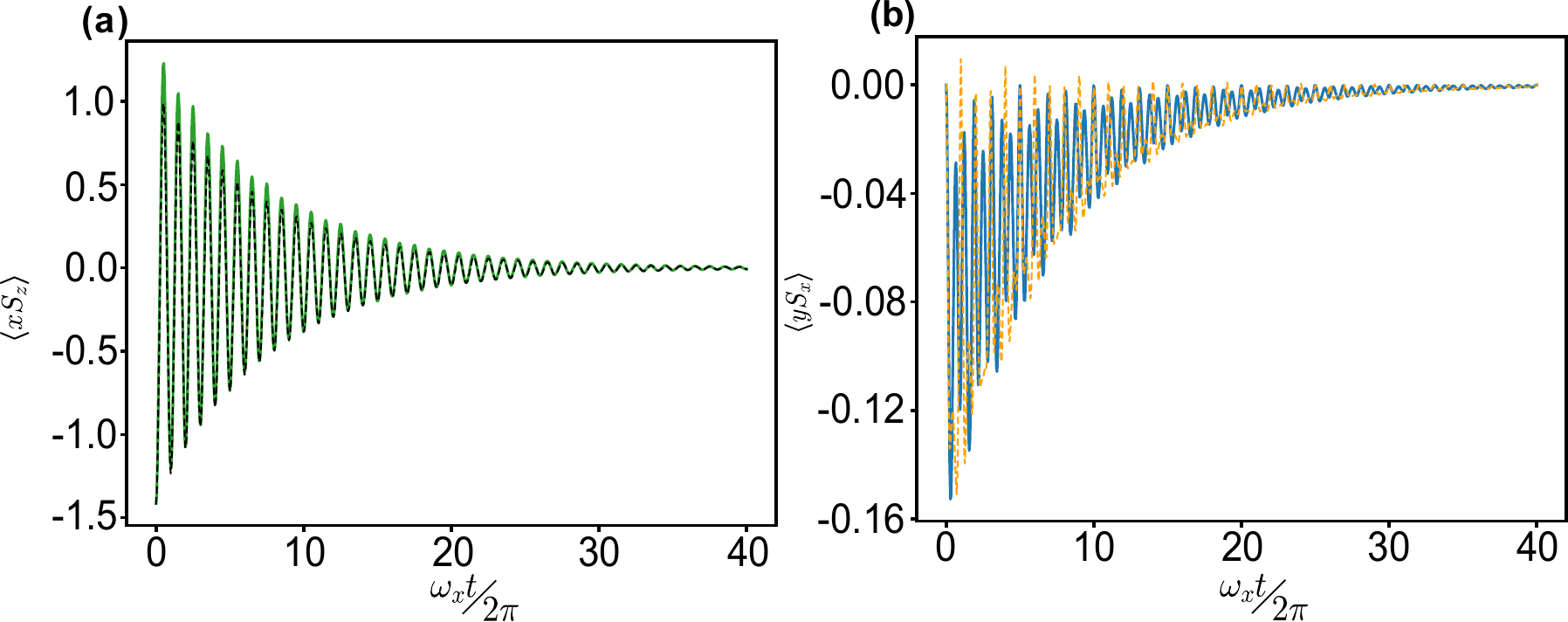}
	\caption{Correlation 
 between spin and phonon modes in the dissipation regime. (\textbf{a}) $\langle xS_z\rangle$ (\textbf{b}) $\langle yS_x\rangle$. 
		Here $(G_x,G_y)=2\pi\times (0.22,0.86)$ MHz (solid line), and~ $(G_x,G_y)=2\pi\times (1,1)$ MHz (dashed line), respectively. Other parameters are $\alpha_x=\sqrt{2}$, $\alpha_y=0$, $\omega_x=2\pi\times 1$ MHz, $\omega_y = 2\pi\times 1.6$ MHz and $\gamma_S = 0.13 \mu s^{-1}$.  
	}%
	\label{fig7}%
\end{figure}
%
The correlations between spin operators associated with the Rydberg levels and phonon also show intriguing properties. We can see that the correlation oscillates initially and~decays with time. Eventually, all the correlations between spins and phonons vanish at a longer time. This is because the left ion is occupied by state $|g\rangle$ when $t\to \infty$. As~a result, the~collective spin sector is gone completely. The~time scale of the damping is largely determined by the time scale of the collective spin operator, i.e.,~lifetime in the state $\vert 0\rangle$.  On~the other hand, the~correlation is robust against changing parameters. By increasing the coupling of $G_x$ and $G_y$, the~oscillation amplitude changes~insignificantly.

\subsection{Dynamics of Individual Rydberg~Ions}
\begin{figure}[H]%
	\includegraphics[width=\linewidth]{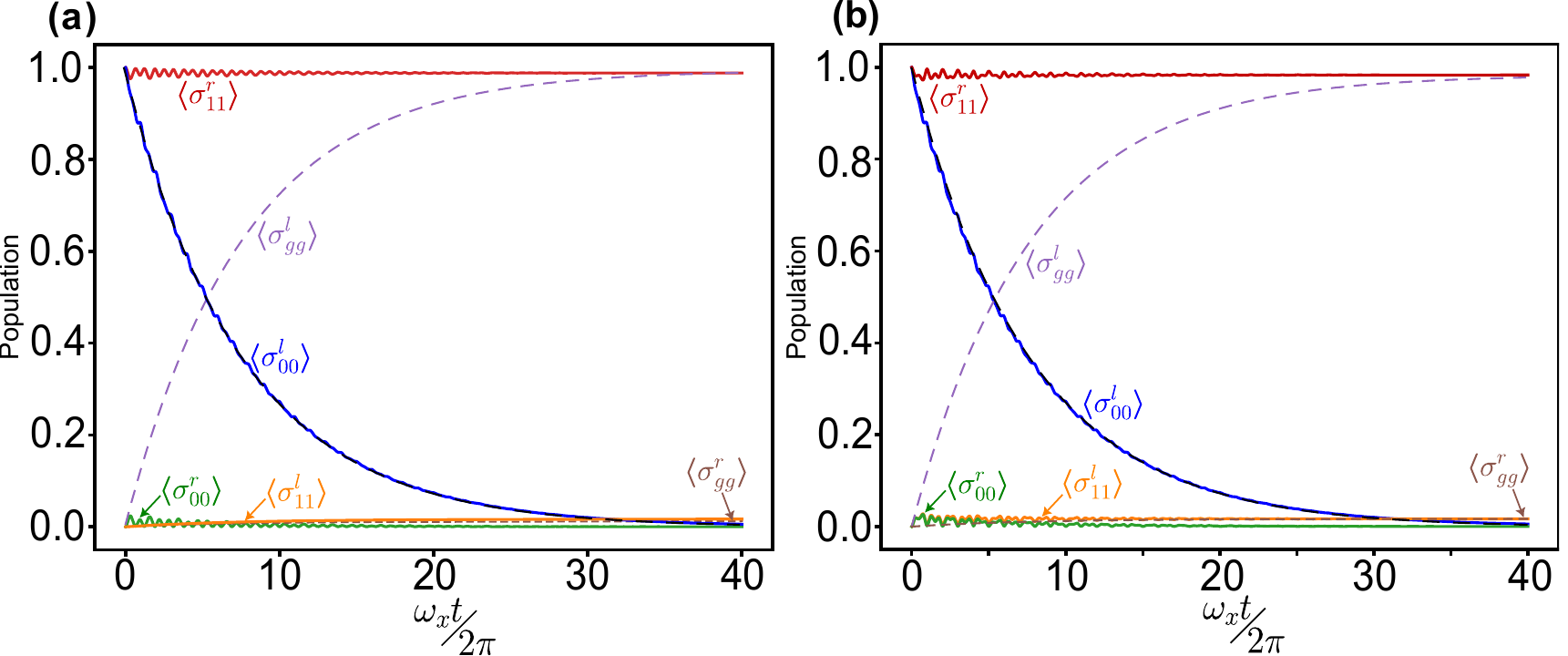}
	\caption{Evolution of individual Rydberg ions with the coupling (\textbf{a}) $(G_x,G_y)=(0.22,0.86)$, (\textbf{b})~$(G_x,G_y)=(1,1)$ for the decay rate $\gamma_S = 0.13 \mu s^{-1}$. The~population of the level $\langle \sigma^l_{gg}\rangle$ increases due to the spontaneous decay from the level $\langle \sigma^l_{00}\rangle$. The~population of the Rydberg state $\langle \sigma^l_{00}\rangle$ follows an exponential decay $e^{-\gamma_{S} t}$ represented by the dotted black line. 
	Population dynamics of the right ion remains unaffected, as~the decay of state $\vert 1\rangle$ is not taken into account due to its very long lifetime. }%
	\label{fig6}%
\end{figure}

We now investigate dynamics of individual ions, which allows us to understand how state $\vert 0\rangle$ decays to state $\vert g\rangle$, and~hence results in our understanding of the time scale. In~Figure~\ref{fig6}, we show the time variation in the population of various electronic states.
For the coupling case $G_x= 2\pi \times 0.22$ MHz and  $G_y= 2\pi \times 0.86$ MHz, it is evident from Figure~\ref{fig6}a that the population in state $|0\rangle$ of the left ion decreases with time. It happens as the Rydberg state $|0\rangle$ decays to the ground state $|g\rangle$ and the population of the ground state $|g\rangle$ increases. At~later times, the~occupancy of the state $|0\rangle$ of the left ion is zero, while it is fully occupied by the $|g\rangle$ state. The~occupancy of the right ion remains in the level $|1\rangle$, as there is no spontaneous decay from this level. Instead, we observe weak oscillations due to the spin--phonon~coupling. 

This simulation shows that the left ion completely decays to the state $|g\rangle$. As~a result, the~collective spin sector defined by $S_\xi$ does not exist anymore. As~the CI is formed in the collective spin sector, this means that the effect of CI decays due to the spontaneous emission of the Rydberg state. However, when the coupling is as strong as it is for the coupling case $G_x = G_y= 2\pi \times 1$ MHz shown in Figure~\ref{fig6}b, we see oscillatory behaviors of the single spin quantities initially before~the Rydberg state of the left ion decays to the ground state. Compared to the dynamics of the collective spin sectors, the~oscillation amplitude is relatively small here. This suggests that it would be easier to observe the oscillatory CI dynamics through measuring the collective spin~states.

Note that this simulation has neglected the decay from the Rydberg state $|1\rangle$. Including the decay of the state $|1\rangle$, time scales of the collective spin sector will not change dramatically, as~state $\vert 1\rangle$ will decay in a time scale given by $1/\gamma_1$.  As~$\gamma\gg \gamma_1$, such decay will be much slower than the decay of the state $|0\rangle$. As~we are interested in the CI dynamics, the decay of the state $|1\rangle$ thus can be neglected in this~study.


%

%

\section{\label{sec5}Summary and~Conclusions}
We have studied dissipative spin--phonon dynamics near a CI. The~CI is engineered in a system of two trapped Rydberg ions. The~spins and phonons are coupled through the long-range dipole--dipole interaction between the two Rydberg ions. We have considered experimentally relevant situations, where the highly excited Rydberg states spontaneously decay to the low-lying ground states~\cite{glukhovLifetimesRydbergStates2013}. Using a quantum master equation, we have provided case studies to show that the interplay between the dissipation and spin--phonon coupling influences the phonon and spin dynamics. When the spin--phonon coupling is strong, it is found that oscillatory dynamics, due to the spin--phonon coupling, can be observed before the collective spin sector decays completely. The~time scale is largely determined by lifetimes of the Rydberg state with a larger decay rate (i.e., $|0\rangle$). Though~Rydberg states have finite lifetimes, this study shows that the CI dynamics can be observed before Rydberg states decay to the other low-lying electronic states. This work has focused on the effect due to finite Rydberg lifetimes. Other factors, such as micromotions~\cite{mullerTrappedRydbergIons2008,martinsImpactMicromotionExcitation2024} and motional heating~\cite{PhysRevA.61.063418}, affect the dynamics of trapped ions in general. How the dynamics near a conical intersection are affected by these factors is worth investigating in the future. Such a study is of relevance to the current Rydberg ion~experiments.



\authorcontributions{Conceptualization, W.L.; methodology, W.L. and R.N.; software, M.C. and W.L.; validation, W.L., R.N. and M.C.; formal analysis, M.C. and W.L.; investigation, M.C., R.N. and W.L.; resources, M.C.; data curation, M.C. and W.L.; writing---original draft preparation, M.C. and W.L.; writing---review and editing, M.C., R.N. and W.L.; visualization, W.L.; supervision, W.L. and R.N.; project administration, W.L.; funding acquisition, W.L. and R.N. All authors have read and agreed to the published version of the~manuscript.}

\funding{W.L. acknowledges support from the EPSRC through Grant No.~EP/W015641/1, the~Going Global Partnerships Programme of the British Council (Contract No.~IND/CONT/G/22-23/26), and~the International Research Collaboration Fund of the University of Nottingham. M.C. acknowledges the funding support from the Department of Science and Technology, Government of India, through the I-HUB Quantum Technology Foundation, IISER Pune, India and the FWO and the F.R.S.-FNRS as part of the Excellence of Science program (EOS project 40007526) at University of Liège, Belgium. R.N. acknowledges DST-SERB for the Swarnajayanti fellowship (File No. SB/SJF/2020-21/19) and MATRICS Grant No.
MTR/2022/000454 from SERB, Government of~India.}

\institutionalreview{Not applicable.}

\informedconsent{Not applicable.}

\dataavailability{The numerical data that support the plots within this paper and other findings of this study are available from the authors upon reasonable request.}

\conflictsofinterest{The authors declare no conflicts of interest. The~funders had no role in the design of the study; in the collection, analyses, or~interpretation of data; in the writing of the manuscript; or in the decision to publish the~results.} 





\appendixtitles{yes} 
\appendixstart
\appendix

\section[\appendixname~\thesection]{Parity Invariance of the Physical Hamiltonian} 
\label{App A}
Here, we show the parity symmetry associated with the Hamiltonian (\ref{interham_oper}) such that the total excitation number is a~constant.

Using the bosonic and spin commutation relations, it can be shown that
\begin{eqnarray}
    e^{i\pi N_y} a_y e^{-i\pi N_y} & = & e^{-i\pi}\sum_{n_y = 0}^{\infty} \sqrt{n_y+1} | n_y\rangle \langle n_y+1 | \nonumber \\ 
    & = & - a_y \\
    S_z S_x S_z & = & - S_x
\end{eqnarray}
and hence, we find that under the parity operation $\mathcal{P} = S_ze^{i\pi N_y} $, the~bosonic and spin operators transform into $\mathcal{P} a_{\xi} \mathcal{P}^\dagger = \pm a_{\xi} $ where the minus sign appears in the $\xi$-phonon mode along the $y$ direction only and $\mathcal{P} S_x \mathcal{P}^\dagger = -S_x$, 
 $\mathcal{P} S_z \mathcal{P}^\dagger = S_z $ due to the commutative property.
It proves that the parity transformation leaves the Hamiltonian invariant, i.e.,
\begin{eqnarray}
    \mathcal{P} H \mathcal{P}^\dagger = H
\end{eqnarray}

\section[\appendixname~\thesection]{Mean Field Equations}
\label{App B}
We employ the mean field (MF) analysis to explore the steady state of the system in the long time limit. 
The time evolution of mean values $\langle O\rangle$ of an operator $O$ is  calculated using the semiclassical approximation by neglecting two-body correlations~\cite{tomadinNonequilibriumPhaseDiagram2011}.
%
%
%
%
Within this approximation, we obtain equations of motion for various expectation values of operators,  
\begin{eqnarray}
	\Dot{A}   &=&  -i  \omega_x A -i G_x  \langle S_z \rangle, \nonumber
	\\
	\Dot{B}  &=& -i  \omega_y B -i G_y \langle S_x \rangle, \nonumber\\
	\langle \Dot{S}_x \rangle  &=& -2 G_x \langle S_y \rangle (A+A^*) - \frac{\gamma_S}{2} \langle S_x \rangle, \nonumber\\
	\langle \Dot{S}_y \rangle  &=& 2 G_x \langle S_x \rangle (A+A^*) - \frac{\gamma_S}{2} \langle S_y \rangle, \nonumber\\
	\langle \Dot{S}_z \rangle  &=& 2 G_y \langle S_y \rangle (B+B^*) - \gamma_S \langle S_z \rangle,\nonumber
\end{eqnarray}
%
where we have defined $A = \langle a_x \rangle$ and $B=\langle a_y \rangle$.
The solution for the steady state is obtained by setting the derivative equal to zero. We find the simple steady state solutions, where $A=B=0$ and $\langle S_x \rangle = \langle S_y \rangle = \langle S_z \rangle = 0 $, i.e.,~the state is obtained with the vanishing expectation values. This is not surprising, as the whole population will be occupied by the state $|g\rangle$ at long times. Such results also hint to us that we should focus on the transient dynamics instead of the steady state. Due to the strong phonon and spin coupling, the~mean field approach will not be valid in the exploration of the dynamics, which is demonstrated by the numerical simulation in the main~text.

\begin{adjustwidth}{-\extralength}{0cm}

\reftitle{References}

\PublishersNote{}
\end{adjustwidth}
\end{document}